\documentclass[reprint,pre,amssymb,amsmath,superscriptaddress,aps, floatfix]{revtex4-2}
\usepackage{amsmath,amsthm,amssymb,amsfonts}
\usepackage{graphicx,color}
\usepackage[outdir=./Figs/]{epstopdf}
\usepackage{siunitx}
\usepackage{hyperref}%
\usepackage{xfrac}
\usepackage{xcolor}
\usepackage{ulem}
\usepackage[T1]{fontenc}
\usepackage{lineno}
\usepackage{multirow}
\usepackage[version=4]{mhchem}
\expandafter\ifx\csname package@font\endcsname\relax\else
 \expandafter\expandafter
 \expandafter\usepackage
 \expandafter\expandafter
 \expandafter{\csname package@font\endcsname}%
\fi
\begin{document}
\title{The role of tumbling in bacterial scattering at convex obstacles}

\author{Theresa Jakuszeit}
\email{theresa.jakuszeit@curie.fr}
\affiliation{Institut Curie, PSL Research University, CNRS UMR168, 75005 Paris, France}

\author{Ottavio A. Croze}
\affiliation{School of Mathematics, Statistics and Physics, Newcastle University, Newcastle upon Tyne NE1 
7RU, United Kingdom }
\date{\today}

\begin{abstract}
Active propulsion, as performed by bacteria and Janus particles, in combination with hydrodynamic interaction results in the accumulation of bacteria at a flat wall. However, in microfluidic devices with cylindrical pillars of sufficiently small radius, self-propelled particles can slide along and scatter off the surface of a pillar, without becoming trapped over long times. This non-equilibrium scattering process has been predicted to result in large diffusivities, even at high obstacle density, unlike particles that undergo classical specular reflection. Here, we test this prediction by experimentally studying the non-equilibrium scattering of pusher-like swimmers in microfluidic obstacle lattices. To explore the role of tumbles in the scattering process, we microscopically tracked wild-type (run and tumble) and smooth-swimming (run only) mutants of the bacterium \textit{Escherichia coli} scattering off microfluidic pillars. We quantified key scattering parameters and related them to previously proposed models that included a prediction for the diffusivity, discussing their relevance. Finally, we discuss potential interpretations of the role of tumbles in the scattering process and connect our work to the broader study of swimmers in porous media.
\end{abstract}

\flushbottom
\maketitle

\section{Introduction}


Both in the laboratory and the natural environment, active particles regularly encounter physical boundaries: synthetic microswimmers can be guided in microfluidic channels \cite{Volpe2011}, sperm cells follow the female reproductive tract to reach the egg cell \cite{Elgeti2010}, and immune cells must navigate the extracellular matrix to respond to danger signals \cite{Moreau2018}.
Accumulation at boundaries may even be essential for biological function, e.g. at the onset of biofilm formation by bacteria \cite{Drescher2011, Secchi2020}. Boundary-induced accumulation is indeed a hallmark of active particles, and the characteristics of a persistent self-propoulsion are sufficient to explain this phenomenon \cite{Elgeti2015, Bianchi2017}. 

Once at a boundary, however, the characteristics of the surface interaction differ between the various types of active particles, depending on, for example, flow fields surrounding the particle body or geometrically determined steric effects. An important class of active particles are microswimmers, such as bacteria and microalgae, which are commonly distinguished based on their flow profile as pushers and pullers, respectively \cite{Elgeti2015review, Lauga2009}.   
For puller-type particles swimming parallel to a wall, the passive hydrodynamic interaction is generally repulsive, and it has been shown that the model microalga \textit{C. reinhardtii} is reorientated at a boundary due to direct contact interactions of its flagella with the surface \cite{Kantsler2013}. 
Pusher-type particles such as bacteria, on the other hand, align their travelling direction upon impact with a wall and escape the surface only at long times \cite{Hu2015}. Once they are trapped by the boundary, the bacteria swim in circles due to the clockwise rotation of the cell body \cite{Lauga2006}. 
To escape the wall again, the bacterium has to reorient sufficiently. A wild-type \textit{E.coli} bacterium has two means to achieve such a reorientation: rotational diffusion or tumbling. Rotational diffusion, with diffusion coefficient $D_r$, is usually a slow process; based on hydrodynamic theory the escape time has been predicted to scale as $\exp (D_r/D_r^*)$, where $D_r^*$ is the rotational diffusion coefficient in the direction perpendicular to the surface \cite{Drescher2011}.
Tumbling is much faster and could be a way of reducing trapping time. However, the tumbling frequency close to a flat surface might be reduced by hydrodynamic effects, which could prevent the unbundling of flagella required for tumbling, as has been demonstrated for \textit{E.coli}  \cite{Molaei2014}. 
While lab surfaces can be flat, those in natural environments may be neither perfectly flat nor smooth. In particular, bacteria have been shown to accumulate at surfaces patterned to have concave curvatures, while convex curvatures may reduce the trapping time \cite{Mok2019, Secchi2020}.
However, even for convex curvatures, there is a radius of curvature above which the particles are trapped, that is, if the curvature of the surface is too small, the boundary resembles a flat wall for bacterium. For smooth-swimming \textit{E.coli}, Sipos {\it et al.} \cite{Sipos2015} determined this trapping radius at 50\textmu m; for radii above this value more than half the cells stayed at an obstacle for more than 3s after collision, which could be explained by a hydrodynamic model. As shown by \cite{Sipos2015} and \cite{Spagnolie2015}, this trapping threshold depends on bacterial properties, in particular its size and dipole strength. 
Below the trapping radius, pusher-type particles scatter at convex surfaces.  Similar to the interaction at a flat wall \cite{Bianchi2017}, both steric and hydrodynamic effects may be involved in the scattering process \cite{Kuron2019, Spagnolie2015, Hoeger2021}.
While some recent studies shed more light on the scattering interaction for a range of different shapes and sizes of obstacles \cite{Dehkharghani2023, Hoeger2021, Krishnamurthi2022, Takaha2023}, it remains unclear how tumbling events might influence the scattering process.
For complex 3D porous media, it was shown that bacteria only escape traps once they reoriented sufficiently, which could be improved by flagellar unbundling \cite{Bhattacharjee2019}. The authors speculated that the geometry of the flagellar arrangement and pore size, as well as hydrodynamic interactions, might influence the ability of a cell to tumble. 
While both smooth-swimming \cite{Takaha2023, Sipos2015} and wild-type bacteria, which perform tumbles \cite{Dehkharghani2023, Krishnamurthi2022, Hoeger2021} have been used for studies involving obstacles, an analysis of the role of tumbles during interactions at convex boundaries is missing to date. This is despite the potential of tumbles as a mechanism to escape obstacles. 

As macroscopic transport arises from the underlying microscopic dynamics, the non-trivial interaction with boundaries can significantly affect the macroscopic behaviour of microswimmers in complex environments, and active matter in confinement has thus been a focus of recent active matter research \cite{Al2022,Bechinger2016}. 
For example, in the classical narrow-escape problem microalgae showed a faster escape than expected for Brownian particles or chaotic Billiards thanks to cell-wall interactions \cite{Souzy2022}.
Interaction with colloids can even increase bacterial propagation due to increases in speed \cite{Kamdar2022} and forward scattering \cite{Makarchuk2019}. These and other studies use wild-type bacteria, which tumble. However, as mentioned above, the role of tumbling on surface scattering has been not previously quantified. 

In this paper, we compare the scattering of wild-type and smooth-swimming \textit{E. coli} in regular obstacle lattices, and use this to evaluate the macroscopic population transport. We first describe the experimental setup as well as the microscopic and macroscopic analysis employed to describe the scattering dynamics. Next, we quantify and compare the details of the scattering behaviour of the smooth-swimming mutant of \textit{E. coli} to its wild-type, to identify the influence of tumbles. We find that tumbling can significantly reduce the time spent at an obstacle for cells approaching it with a large collision angle. Finally, we discuss the diffusive transport that results from the scattering of populations of the two strains.

\section{Methods}

\subsection{Experimental setup}

Experiments were preformed using \textit{E. coli} strains AD52 (AB1157 motility wild-type \cite{Dewitt1962} with plasmid expressing eGFP pWR21) and AD83 (an AB1157 $\Delta$cheY smooth-swimming mutant, JSL1 \cite{Schwarz2016}, with plasmid expressing eGFP pWR21). The preparation of the bacterial cultures followed standard protocol developed by H.C. Berg for motility studies using \textit{E.coli} as outlined, e.g., in \cite{Schwarz2016}. In short, cultures were grown from frozen stocks on Luria Broth agar plates ($10$g/L tryptone, $5$g/L yeast extract, $10$g/L \ce{NaCl}, $1.5$g/$100$mL agar) overnight at \SI{30}{\celsius} (New Brunswick Scientific, Inova 42 R). A single culture was transferred from plates to liquid LB medium and grown overnight in LB at \SI{30}{\celsius} and $200$rpm. Cultures were then diluted 1:100 in Tryptone broth ($10$g/L tryptone, $10$g/L \ce{NaCl}) and incubated at \SI{30}{\celsius} and $200$rpm for $4$h (up to $\mathrm{OD}_{600} \sim 0.4-0.5$). The growth media were supplemented with $100\mu$g/ml ampicillin and $30\mu$g/ml kanamycin, where needed, e.g. to retain plasmids. At the end of the second growth phase, 1mL of culture was washed three times by centrifuging at $8000$g at \SI{20}{\celsius} for 2min, discarding the supernatant and resuspending the pellet gently before adding 1mL of Berg's motility buffer (BMB: $6.2$mM \ce{K2HPO4}, $3.8$mM \ce{KH2PO4}, $67$mM \ce{NaCl}, $0.1$mM \ce{EDTA}). After the final centrifugation, the pellet was resuspended in BMB +$4\%$ of bovine serum albumin (BSA) to prevent the surface attachment of bacteria.

Polydimethylsiloxane (PDMS) devices were fabricated
according standard soft lithography techniques, using a 1:10 mixture of elastomer:silicone (SYLGARD 184). The devices were cured at \SI{60}{\celsius} for 2h. After atmospheric plasma treatment for 10s (diener Femto plasma system), the PDMS devices were bonded to glass coverslips and stored at \SI{60}{\celsius} for 15min to improve bonding. The resulting channels had a height of $50\mu$m and were filled with hexagonal lattices of pillars with varying radius and distance (centre-to-centre separation). The pillars have a radius of either $R=16\mu$m or $R=36\mu$m, which is well below the critical trapping limit $R\sim50\mu$m \cite{Sipos2015}. 
The microfluidic channel was imaged on an Olympus IX73 Inverted Microscope using fluorescence imaging (Prior Lumen 200 illumination) at 10$\times$ magnification. Image sequences were acquired at $20$fps using a CMOS camera (Grasshopper3 GS3-U3-23S6M, $1.71\text{pixel}/\mu$m). The contrast was enhanced in MATLAB by stretching the pixel values based on the standard deviation of the image. Finally, a bandpass filter was applied to enhance edges and reduce low-frequency noise.
Particle tracking was based on the algorithm developed by Crocker and Grier \cite{Crocker1996}, and the obtained trajectories were smoothed using a Gaussian-weighted moving average.
To detect the position of pillars, their outline was determined from a bright field image using a circular Hough transform implemented in MATLAB's \textit{imfindcircles} function. 

\subsection{Scattering analysis}
\begin{figure}
	\includegraphics[width=\columnwidth]{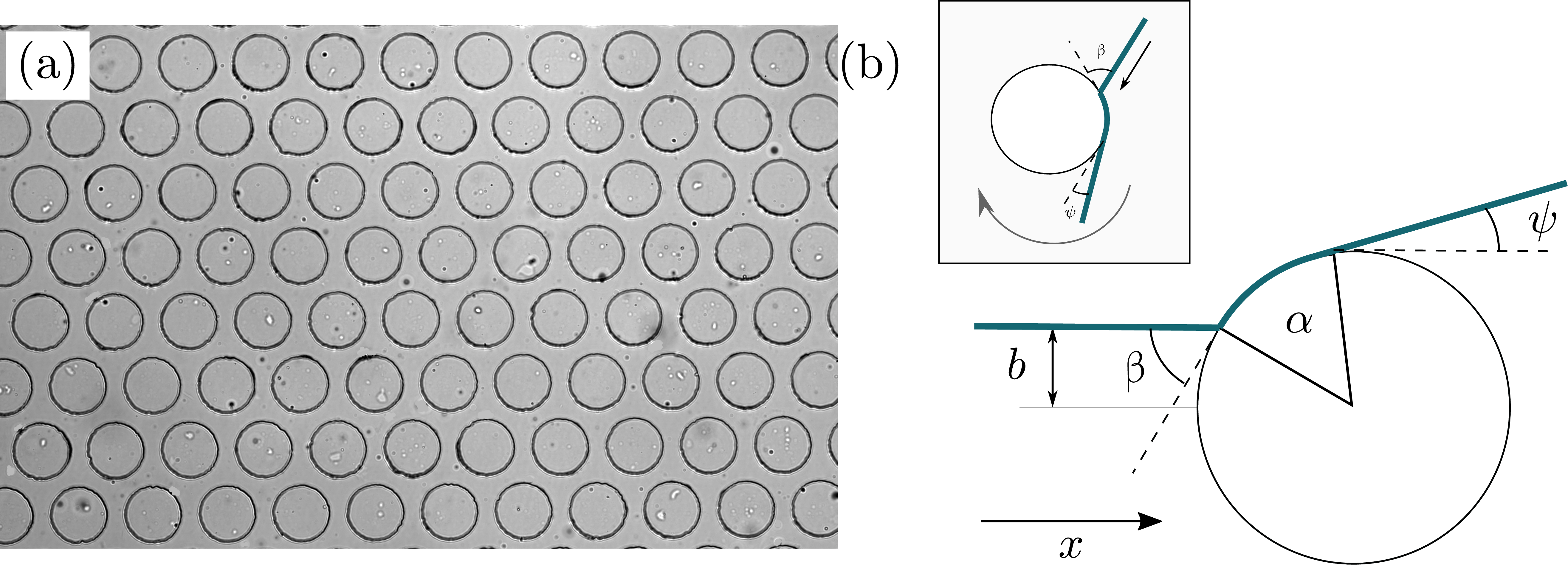}
			\caption{Microfluidic setup and microscopic analysis. (a) Brightfield image of a microfluidic channel with obstacles of radius $R=36\mu$m and distance $d=100\mu$m. This image is used to identify the pillars as circles (b) The impact parameter $b$ is the $y$-component of the rotated trajectory before collision, which relates to the collision angle $\beta$ via $\cos(\beta)=b/R$. The reorientation angle due to obstacle collision is the orientation of the rotated track after collision, and is a combination of the collision angle $\beta$ and polar angle $\alpha$, $\psi=\beta-\alpha$ \cite{Jakuszeit2019} Inset: trajectory before rotation.}
	\label{fig: ImpactParameterDerivation}
\end{figure}


In order to study bacterial interactions with pillars we need a method to identify scattering events, which was done as follows. First, those parts of trajectories were identified which were in contact with an obstacle, i.e. the distance between an obstacle centre and particle position $\mathbf{r}$ satisfied
\begin{equation}
\vert \mathbf{O} - \mathbf{r} \vert < R+ \epsilon,
\label{eq:PillarInteraction}
\end{equation} 
where $\mathbf{O}$ and $R$ are the obstacle centre vector and radius, respectively. The threshold $\epsilon$, which constitutes a layer around the obstacle, was chosen empirically as $1.0\mu$m based on the typical width of \textit{E.coli} cells \cite{Cesar2017}, unless discussed otherwise. 
Once the bacterium-pillar interactions were identified, partial trajectories that correspond to $0.4$s before and after collision were retained.
Subsequently, each identified interacting trajectory was shifted such that the centre of the obstacle was at the origin. Because the same particle might interact with multiple obstacles, trajectories were divided up and each part of the trajectory was shifted separately. Next, making use of the symmetry of the pillars, particle tracks were rotated clockwise based on their orientation before impact such that their incoming direction was aligned with the $x$-direction, see Fig.~\ref{fig: ImpactParameterDerivation}. We then define the impact parameter $b$ as the $y$-component of the rotated track before collision. By virtue of this definition, an impact parameter $b=0$ thus corresponds to a head-on collision. The relationship between the impact parameter $b$ and the collision angle, defined as the angle between the particle orientation and the surface tangent at collision point, $\beta$, is given by the alternate angle theorem as $\cos(\beta)=b/R$, see Fig.~\ref{fig: ImpactParameterDerivation}(b). Hence, a small collision angle $\beta$ corresponds to a large absolute impact parameter $\vert b \vert$ and vice versa. In the following, both parameters will be used interchangeably. 
The reorientation angle due to the obstacle interaction, $\psi$, was determined as the orientation of the track after the particle leaves the obstacle, see Fig.~\ref{fig: ImpactParameterDerivation}(b). 
Finally, the residence time follows for each trajectory simply as the number of frames in which Eq.~\eqref{eq:PillarInteraction} is satisfied divided by the frame rate.

We seek to relate microscopic dynamics with macroscopic diffusive transport, which we can derive from particle trajectories via the mean squared displacement (MSD). For a persistent random walk with speed $v$ and persistent time $\tau$, an analytical expression for the MSD can be derived as \cite{Lovely1975}
\begin{align}
\langle r(t)^2 \rangle =2 v^2 \tau t + 2 v^2 \tau^2\left(e^{-t/\tau}-1\right),
\label{eq:MSDprw}
\end{align}
where the first term on the right hand side corresponds to the long term diffusive behaviour, with diffusivity $D=v^2 \tau/2$, while the second term is caused by the ballistic motion at short times. The cross-over from ballistic to diffusive behaviour is characterised by $\tau$. 
For wild-type \textit{E.coli}, a run-and-tumble particle with a forward bias during tumbles, we also need to define the effective run time
\begin{equation}
\tilde{\tau}=\dfrac{\tau}{1-\langle \cos \psi_T \rangle },
\end{equation}
where the value of $\tau$ is obtained from fitting Eq.~\ref{eq:MSDprw} and $\langle \cos \psi_T \rangle$ is the mean cosine of the tumbling angle $\psi_T$ \cite{Lovely1975}.

\section{Results}
\label{sec:Results}

\begin{figure}
	\includegraphics[width=\columnwidth]{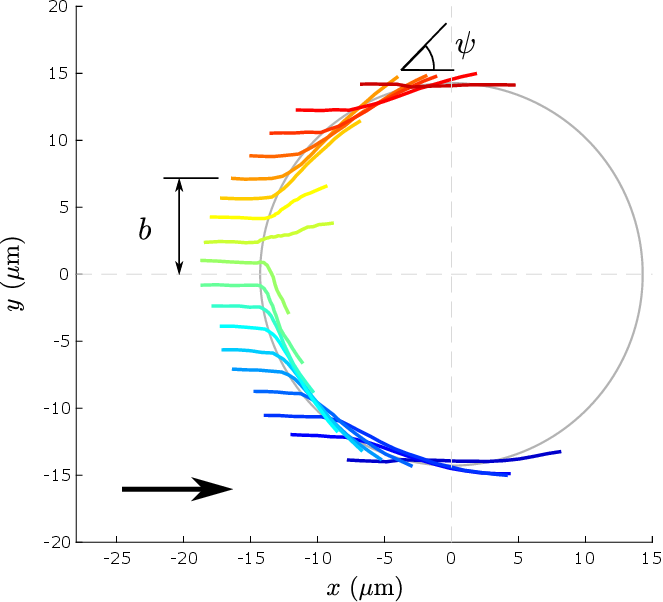}
	\caption{Average trajectory upon obstacle collision. Particle trajectories from the smooth-swimming mutant are rotated such that their incoming direction is aligned with the x-direction. The impact parameter $b$ is the y-component of the rotated track. Trajectories are binned according to impact parameter $b$ and the average trajectory upon obstacle collision is shown given a certain $b$. The collision-induced reorientation, $\psi$, corresponds to the orientation of the rotated track after the obstacle interaction. The trajectories are colour coded based on impact parameter.}
	\label{fig: ImpactParameter}
\end{figure}

We expect that the impact parameter $b$ affects the interaction of the particle with the obstacle. To illustrate this point, particle trajectories were binned according to their impact parameter $b$. We can then obtain the average trajectory given a certain impact parameter as shown in Fig.~\ref{fig: ImpactParameter}, where the trajectories are colour-coded according to their impact parameter.
As can be seen from Fig.~\ref{fig: ImpactParameter}, the magnitude of reorientation varies depending on the impact parameter. Furthermore, the travelling direction around the obstacle is mainly dictated by the sign of the impact parameter. However, the average trajectory of impact parameters close to zero often appears to point into the obstacle, as illustrated in Fig.~\ref{fig: ImpactParameter}. This artefact is caused by a similar proportion of trajectories going in either direction around the obstacle upon collision, and therefore `cancelling' out in the average. This could be explained by either steric or hydrodynamic effects. For example, at small $b$ (i.e. large collision angle $\beta$), rotational diffusion can reorient the cell far enough to seemingly reverse direction \cite{Jakuszeit2019}, while hydrodynamic simulations of squirmers have reported the possibility of `mobility reversals' \cite{Kuron2019}. In the following we will restrict the analysis to the absolute value of the impact parameter scaled by the pillar radius.

\begin{figure}
	\includegraphics[width=\columnwidth]{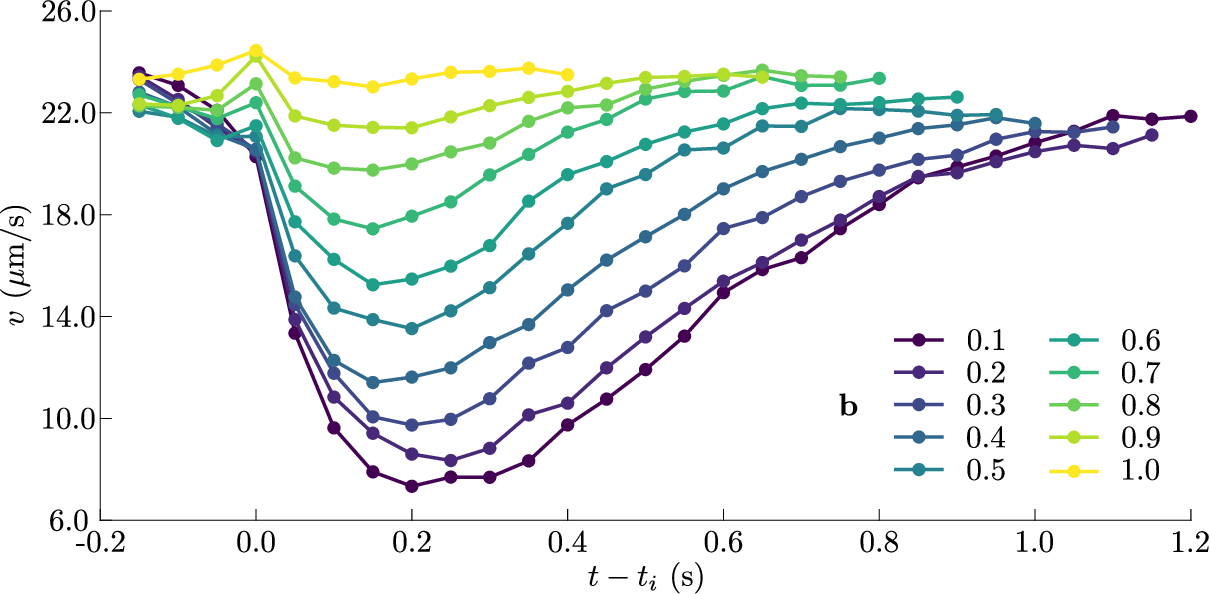}
	\caption{The average instantaneous speed upon obstacle collision depends on impact parameter. Trajectories of smooth-swimming mutant $\Delta$cheY at an obstacle $R=36\mu$m are binned according to impact parameter $b$ and the average instantaneous speed upon obstacle collision at impact time $t_i$ is shown given a certain $b$. The trajectories are colour coded based on impact parameter.}
	\label{fig: ImpactParameterVelocity}
\end{figure}

First, we look at the instantaneous speed during obstacle interaction. As shown in Fig.~\ref{fig: ImpactParameterVelocity}, it decreases abruptly upon impact with the obstacle. The drop in speed was larger for smaller impact parameters and the speed took longer to recover to pre-impact levels. However, the minimum speed was not reached immediately after impact, as would be expected from a purely steric obstacle interaction. 
In addition, the speed already changed before impact. Here we can observe two different types of behaviour: i) cells approaching with a large collision angle (i.e. small $b/R$), slow down, whereas ii) cells approaching with a small collision angle (i.e. large $b/R$) appear to slightly accelerate before impact. In \cite{Bianchi2017}, bacteria with a large collision angle were observed to slow down when approaching a flat wall, which the authors explained by hydrodynamic interactions. Conversely, cells swimming parallel to a flat wall were observed to accelerate in \cite{Khong2021}.

\begin{figure}
	\includegraphics[width=\columnwidth]{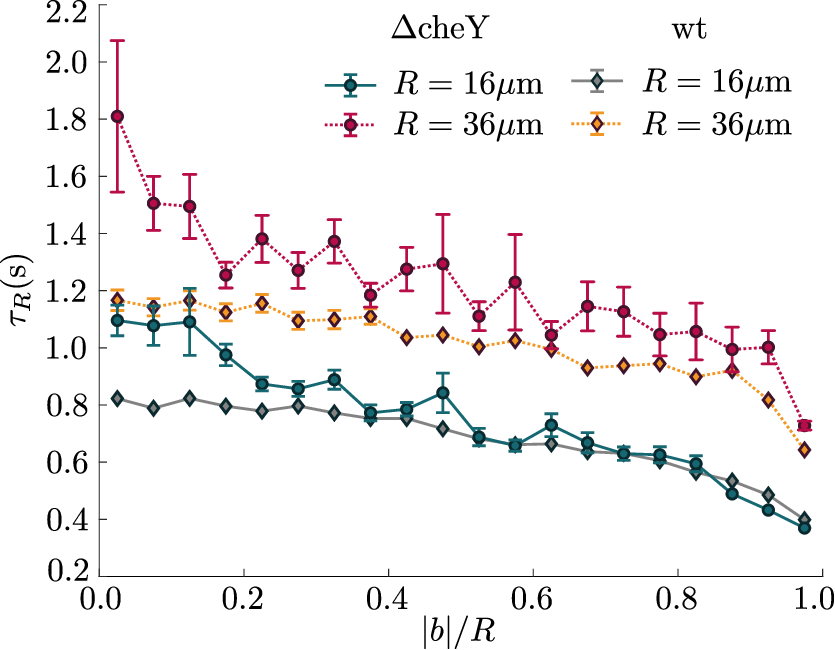}
	\caption{Residence time $\tau_R$ at an obstacle for smooth-swimming mutant $\Delta$cheY (solid lines, circle marker) and wild-type (wt) \textit{E.coli} (dashed lines, diamond marker). The residence time is larger for a larger obstacle. Tumbling reduces the time spent at an obstacle for small impact parameter $\vert b \vert$ for both obstacle radii. The error bars correspond to the standard error over all included samples (bars not shown if smaller than marker). 
}
	\label{fig: ResidenceTimeTumbling}
\end{figure}

The decrease in instantaneous speed during obstacle interaction also has a stark effect on the time that bacteria spent at the obstacle. As shown in Fig~\ref{fig: ResidenceTimeTumbling}, the residence time can be almost twice as large for the smallest impact parameter as it is for the largest $|b|/R$. A larger obstacle radius, which corresponds to a smaller curvature, increases the residence time significantly. The residence time also highlights an interesting difference between the smooth-swimming mutant $\Delta$cheY and wild-type \textit{E. coli}. Whilst the residence time of the wild-type is smaller than the smooth-swimming mutant for small impact parameters, the difference vanishes for large $\vert b \vert$, in particular for the smaller obstacle radius.
The motility pattern of the wild-type differs from the smooth-swimming mutant by the occurrence of tumbling events. Tumbling suppression by as much as $50\%$ has been demonstrated for \textit{E. coli} close to flat surfaces \cite{Molaei2014}, but, even in this case, whenever a tumble occurs, it could also be a means of escape, which reduces the residence time $\tau_R$. For large impact parameters, the orientation of the cell is already close to aligned with the surface and tumbling might not influence the residence time much. Conversely, the influence of tumbles is large at small $b$, where they may remove the cell from the obstacle. 
\begin{figure}
	\includegraphics[width=\columnwidth]{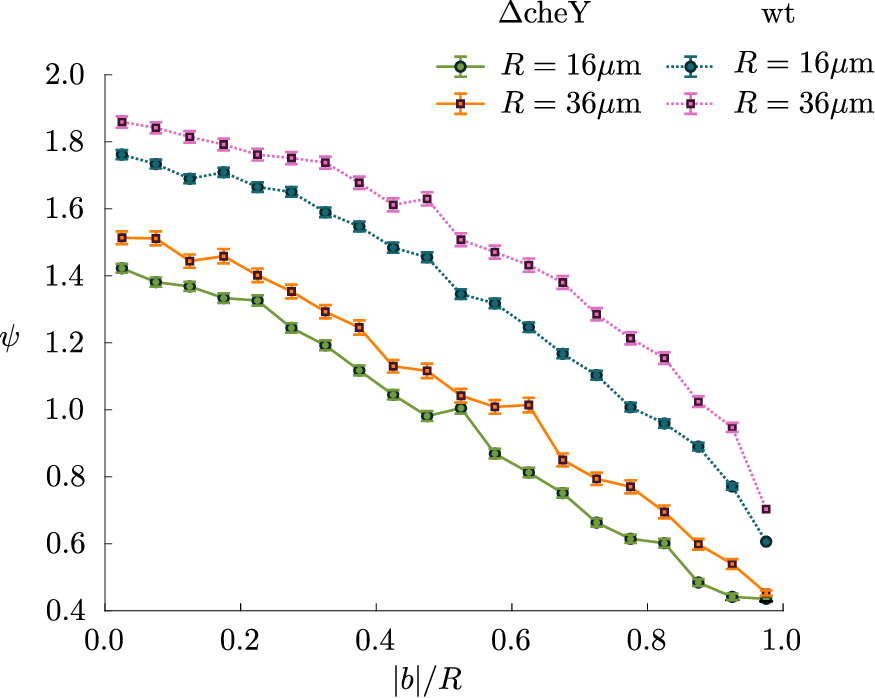}
	\caption{Reorientation depends on impact parameter and obstacle radius for smooth-swimming mutant $\Delta$cheY (solid lines, circle marker) and wild-type (wt) \textit{E.coli} (dashed lines, diamond marker). The reorientation angle $\psi$ decreases with increasing impact parameter $\vert b \vert$. The error bars correspond to the standard error of the sample (bars not shown if smaller than marker). 
}
	\label{fig: ReorientationPsi}
\end{figure}

\begin{figure}
	\includegraphics[width=\columnwidth]{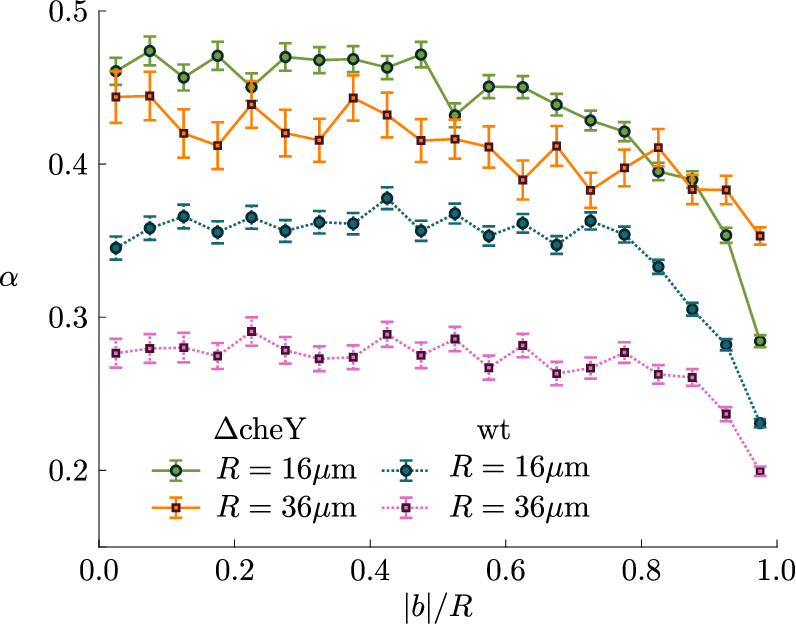}
	\caption{Polar angle $\alpha$ at an obstacle for smooth-swimming mutant $\Delta$cheY (solid lines, circle marker) and wild-type (wt) \textit{E.coli} (dashed lines, diamond marker). Following the picture in Fig.~\ref{fig: ImpactParameterDerivation}, the reorientation angle $\psi$ is a combination of reorientation due collision angle $\beta$ and polar angle $\alpha$. The polar angle $\alpha$ is calculated for individual trajectories according to Eq.~\eqref{eq: ExpAlphaProp}. There is only a weak dependence of $\alpha$ on $\vert b \vert$ but a smaller radius increases $\alpha$. The error bars correspond to the standard error of the sample (bars not shown if smaller than marker). 
}
	\label{fig: ReorientationAlpha}
\end{figure}

Fig.~\ref{fig: ReorientationPsi} shows the reorientation angle $\psi$ as a function of the (absolute) impact parameter rescaled by the obstacle radius $R$.
The reorientation $\psi$ decreases with increasing impact parameter, and a larger pillar radius $R$ leads to a larger reorientation. When a particle collides with an obstacle, it needs to be reoriented at least by the collision angle $\beta$ in order to point away from the surface and escape. Since $\beta$ relates to $b$ and $R$ via $\cos (\beta)=b/R$, the $b$ and $R$ dependence observed in Fig.~\ref{fig: ReorientationPsi} is unsurprising. However, $\psi$ might exceed the expected minimum reorientation $\beta$ since rotational noise or hydrodynamic attraction to the obstacle might increase the reorientation, and could, thereby, introduce further dependence on $b$ and/or $R$. We note that the reorientation angle $\psi$ can be defined as a combination of the collision angle $\beta$ and a polar angle denoted as $\alpha$, if we assume that the particle leaves at a tangent to the obstacle surface,
\begin{equation}
\psi=\beta-\alpha=\cos^{-1} \left( \dfrac{b}{R} \right) - \alpha.
\label{eq: ExpAlphaTheory}
\end{equation}
The polar angle $\alpha$ can, hence, give an indication of a reorientation that exceeds the expected reorientation due to the collision angle $\beta$. As both incident and leaving points on the obstacle are known, we can calculate the polar angle independently as the angle between two points of distance $c$ on a circle of radius $R$ 
\begin{equation}
\alpha=2 \sin^{-1} \left( \dfrac{c}{2R} \right).
\label{eq: ExpAlphaProp}
\end{equation} 
Fig.~\ref{fig: ReorientationAlpha} shows that $\alpha$ is constant for a large range of $|b|/R$, but decreases at large $|b|/R$. The smooth-swimming mutant does not show significant $R$-dependence in $\alpha$, while the wild-type has a smaller $\alpha$ for the larger obstacle radius. As the arc length along the obstacle is defined as $L=\alpha R$, this means that the smooth-swimmer covers a larger distance on the obstacle for a larger obstacle $2$ than a smaller one $1$: $L_2/L_1\approx R_2/R_1$, as $\alpha\approx$ constant. The wild-type will also travel further on a larger obstacle, however here the ratio of $L_2/L_1$ is reduced because of the dependence of $\alpha$ on $R$. This appears compatible with the reduction in residence time for the wild-type.

\subsection{Diffusion in a lattice}

As we do not observe any trapping at single-cell level, the macroscopic transport in pillar lattices is expected to be of diffusive nature \cite{Chepizhko2013}.
If we consider reorientations due to obstacle interactions as effective tumbles, we can use established results for the diffusion coefficient of run-and-tumble particles (RTPs) \cite{Jakuszeit2019} and calculate the effective diffusion coefficient for the smooth-swimming \textit{E.coli} $\Delta$cheY as 
\begin{equation}
\tilde{D}=\dfrac{\tilde{v}^2}{2\left[ D_R + (1- \langle \cos \psi \rangle)/\tau \right] },
\label{eq:DeffcheY}
\end{equation}
where $\tilde{v}$ is the effective swimming speed (defined in Eq.~\eqref{Eq: effective speed}), $D_R$ is the rotational diffusion coefficient and $\tau$ is the reorientation time scale. The rotational diffusion coefficient was estimated using particle tracks from the lattice with the smallest $R/d$ (low obstacle density). The mean squared angular deviation was fitted according to $\langle \varphi(t)^2 \rangle=2D_R\,t$ at long times to obtain $D_R=0.16\mathrm{rad}^2/$s. The mean cosine $\langle \cos \psi \rangle$ was evaluated from our scattering data.

In addition to the obstacle induced `tumbles', an estimate for the wild-type \textit{E.coli} needs to consider biologically-induced tumbles in the RTP model. Assuming these two different types of tumbles represent two independent Poisson processes, the diffusivity of the wild-type can be written as
\begin{equation}
\tilde{D}=\dfrac{\tilde{v}^2}{2\left[ D_R + (1- \langle \cos \psi \rangle )/\tau +(1- \langle \cos \psi_T \rangle ) \mu  \right] },
\label{eq:DeffWT}
\end{equation}
where $\mu$ is the tumbling rate of the bacterium, which was approximated using the tumbling rate at low density, $\mu=0.63\mathrm{s}^{-1}$. The value of the mean cosine for tumbles was taken from \cite{Berg1972} as $\langle \cos \psi_T \rangle=0.35$ ($\psi_T=69^\circ$), and it was assumed that $D_R$ is the same as the smooth-swimmer.

\begin{figure}
	\centering
	\includegraphics[width=0.8\columnwidth]{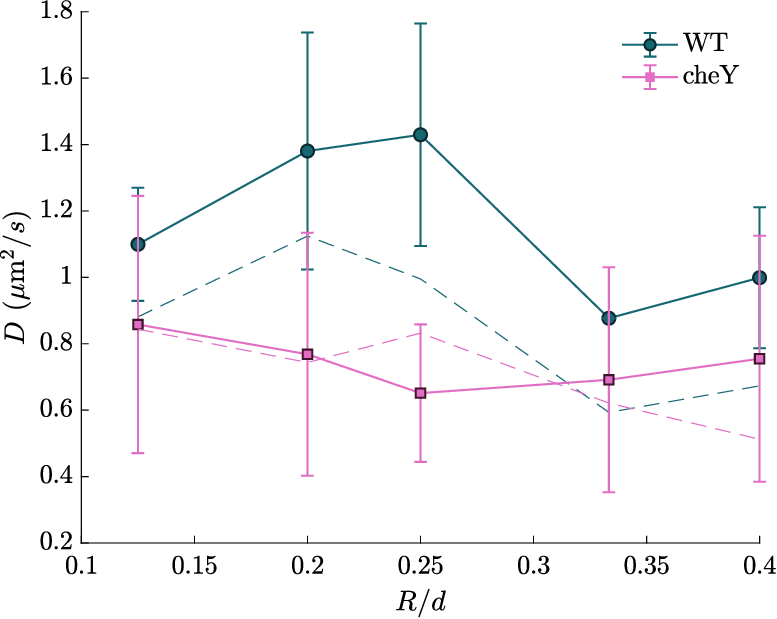}
	 
	\includegraphics[width=0.8\columnwidth]{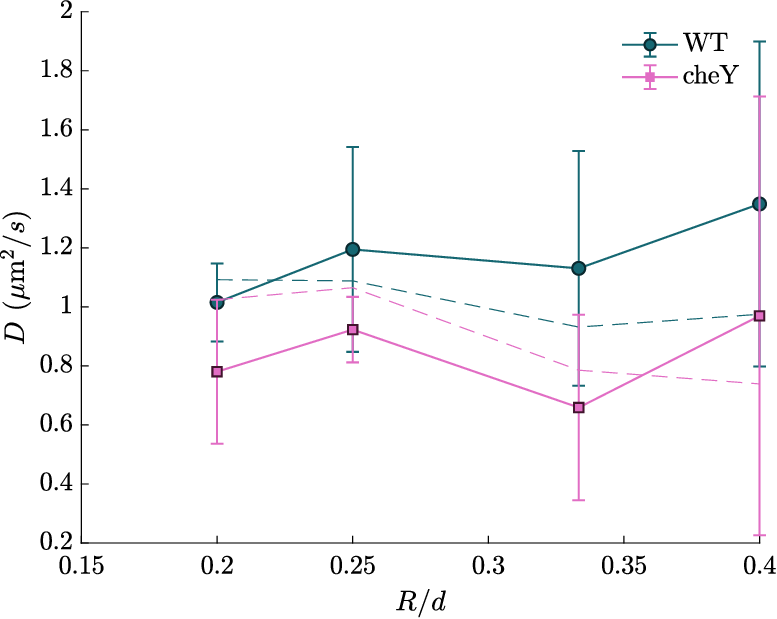}
	\caption{Relative diffusion coefficient for wild-type \textit{E.coli} and smooth-swimming mutant $\Delta$cheY at different obstacle separations for obstacle radius (a) $R=16\mu$ and (b) $R=36\mu$. The diffusion coefficient obtained by fitting the MSD with Eq.~\eqref{eq:MSDprw} is rescaled by the expected free diffusion coefficient $D_0=351\mu \mathrm{m}^2/s$ (WT) and $D_0=1250\mu \mathrm{m}^2/s$ ($\Delta$cheY), see main text.
	The dashed line is the theoretical prediction based on Eq.~\eqref{eq:DeffWT} and Eq.~\eqref{eq:DeffcheY}, respectively. For both strains, the diffusive transport is large even for dense obstacle lattices. The error bars correspond to the standard deviation of different repeats.
	}
	\label{fig: DiffCoeff}
\end{figure}

\begin{figure}
	\centering
 
	\includegraphics[width=0.8\columnwidth]{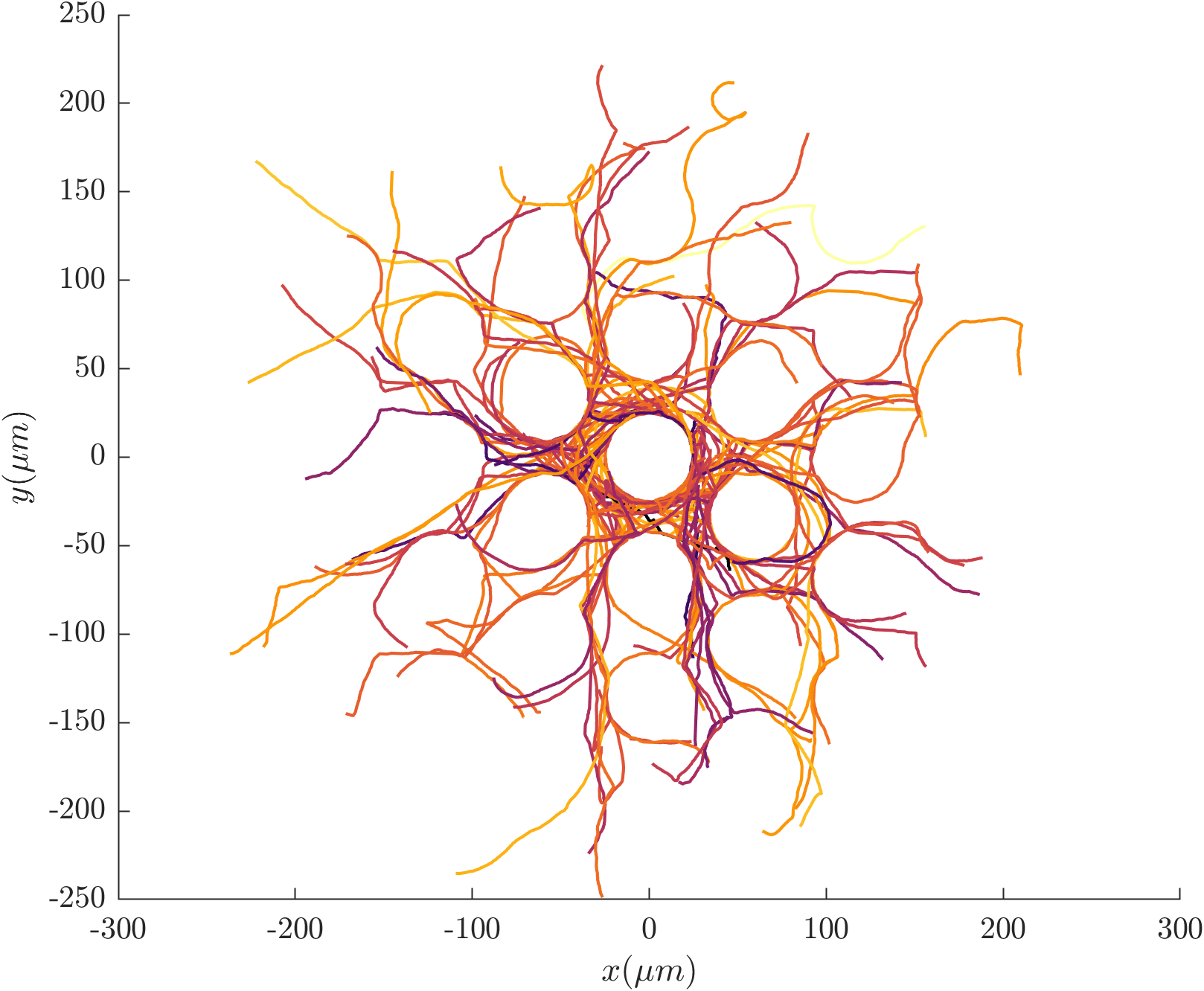}
	\caption{Effect of channels on particle trajectories. 
	Particle trajectories centred based on the closest obstacle at the beginning of the particle trajectory, whose centre constitues the origin. 
	Trajectories of smooth-swimming mutant in an obstacle lattice of $R=36\mu$m and $d=100\mu$m with a fixed duration ($15$s).}
	\label{fig: Channel_effect}
\end{figure}

In Fig.~\ref{fig: DiffCoeff} we compare experimental diffusion coefficients, obtained from MSDs using Eq.~\eqref{eq:MSDprw}, with the above theoretical estimates. The experimental mean diffusivities are large even in dense obstacle lattices, and significantly above the theoretical prediction for both smooth-swimming mutant and wild-type \textit{E.coli}. Examining individual particle tracks, see Fig.~\ref{fig: Channel_effect}, some bacteria appear to follow a `channel' defined by the obstacles for a significant time, as was predicted in \cite{Jakuszeit2019}. The latter study pointed to channeling as a mechanism to explain high diffusivities in dense lattices, which were not captured by the RTP model. Therefore, future work, both theoretical and experimental, should aim to study geometrical guiding effects in the dense obstacle lattices. 

\section{Discussion}


We have studied the interaction of bacteria with cylindrical obstacles in a microfluidic lattice. This simple geometry provides a model system for the quantification of active matter in complex and crowded environments, whose importance continues to increase \cite{Bechinger2016}. While there have been multiple studies of bacteria in complex environments using both tumbling wild-types \cite{Dehkharghani2023, Krishnamurthi2022, Hoeger2021} and smooth-swimming mutants \cite{Takaha2023, Sipos2015}, the effect of tumbles has been thus far neglected. By directly comparing tumbling wild-type \textit{E.coli} and smooth-swimming mutants interacting with an obstacle lattice, we here have shown that tumbling can significantly change the boundary interactions. 

In particular, we found that the difference in residence time between wild-type and smooth-swimmer mutant was most pronounced at small impact parameters, while it diminished at large impact parameters, when the cells only slide past the obstacle. As a tumble on average results in a sudden large reorientation \cite{Berg1972}, its effect might be more pronounced at small impact parameters because, in this case, the reorientation required to escape the boundary is large. In addition to `natural' tumbles, an interesting possibility is that head-on collisions might (additionally) increase the probability of tumbling due to an increased load \cite{Wang2017}. Future experiments with stained flagella could elucidate this aspect further. 
The residence time, in addition, depends strongly on the pillar radius, increasing for larger $R$ values for both strains. A larger radius increases hydrodynamic attraction, as well as the role of noise in escaping the obstacle due to the decreased curvature \cite{Spagnolie2015}. The difference between the residence times for the strains considered is more pronounced for large radii, see Fig.~\ref{fig: ResidenceTimeTumbling}. This could be due to the fact that tumbling may have a stronger influence on the residence time at larger obstacles, because a larger reorientation is required to escape at smaller curvature. As tumbling bacteria statistically spend less time travelling around obstacles with larger radii, this effect could then be the cause for the $R$-dependence in the $\alpha$-values determined for the wild-type in Fig.~\ref{fig: ReorientationAlpha}, compared to the smooth-swimmer, for which $\alpha$ is approximately independent of $R$.

Different studies have underlined the importance of hydrodynamic vs. steric interactions during the scattering process at convex surfaces \cite{Berke2008, Li2009,Bianchi2017,Hoeger2021}. 
In contrast to steric effects, long-range hydrodynamic effects can act at a distance due to the flow field surrounding microswimmers. Notably, theoretical studies predicted an alignment of the swimming direction before direct interaction with the surface resulting from hydrodynamic effects \cite{Spagnolie2015}. Experimentally, microswimmer reorientation during approach has, to the best of our knowledge, not been studied for round obstacles. However, we can make a comparison with flat surfaces, which are locally similar to large obstacles.
A previous study, which released \textit{E. coli} at a defined distance from a flat wall using optical tweezers, did not confirm the theoretical prediction of alignment of the swimming direction prior to hitting the wall \cite{Bianchi2017}. By contrast, in \cite{Khong2021}, it was found that freely swimming \textit{Pseudomonas aeruginosa} and \textit{E.coli} reoriented to be parallel to a flat wall, and that this reorientation was a function of the distance from it, which suggests a hydrodynamic effect. 
Indeed, our results for how the distribution of impact parameters (see Supplementary Material) changes with $\epsilon$ (the interaction layer thickness), also suggest hydrodynamic reorientation prior to interaction with the surface. 
Intriguingly, as shown in Fig.~\ref{fig: ImpactParameterVelocity}, we found that the speed of smooth swimmers decreases for cells approaching an obstacle with a large angle, whereas cells approaching it with a small angle accelerate. We may again compare these results to experimental and theoretical studies at a flat wall. In \cite{Bianchi2017} cells were observed to slow down when approaching a flat surface with a large angle of approach, which was modelled as a result of body-wall hydrodynamic coupling. On the other hand, in \cite{Khong2021} it was found that the average speed increases closer to the wall, where the average orientation of the bacteria changes to be parallel to the wall. The authors explained this observation by a larger increase in the perpendicular to parallel drag coefficient for a rod-like swimmer, as the surface is approached \cite{Katz1974, Katz1975}. Our results for cells approaching a convex surface with a small vs. a large angle thus show similarities with the results for flat walls reported in \cite{Khong2021} and \cite{Bianchi2017}, respectively. Theoretical studies could model these and compare them with our results, those of related investigations \cite{Hoeger2021} and future experiments.The alignment of swimming direction as well as the change in swimming speed before direct impact with the obstacle suggest that hydrodynamics play a role in the system studied here, while a recent study came to the conclusion that hydrodynamic effects are negligible, as a steric model was sufficient to explain the interaction with obstacles below $10$x the body length \cite{Hoeger2021}. However, this conclusion was based on the direct interaction alone and did not take into account any parameters before impact. As the crucial difference between steric and hydrodynamic effects is the ability of the latter to act at a distance, this might explain the inference of this study that the interaction itself may be dominated by steric effects.

Our results for the reorientation $\psi$ shown in Fig.~\ref{fig: ReorientationPsi}, which for the wild-type are qualitatively similar to what was found in \cite{Hoeger2021}, show how for large impact parameters the deflection by a pillar is small, and the difference between different radii vanishes. This can be simply explained by the fact that the interaction is weak when approaching a pillar tangentially, whatever the pillar radius. From the reorientations it was also possible to identify a polar angle $\alpha$, which quantifies reorientation beyond the minimum $\beta$. In \cite{Jakuszeit2019}, it was postulated that swimmers would be re-orientate by a fixed polar angle $\alpha$, providing either sliding (hydrodynamic/steric motivated) of a slide-off (pure steric) boundary conditions. Experimentally, we have shown that this an acceptable approximation for a broad range of impact parameters. At the highest impact parameters, however, $\alpha$ monotonically decreases. Interestingly, for the smooth swimmer $\alpha$ does not show much $R$ dependence, while for the wild-type $\alpha$ is reduced on larger obstacles. As discussed previously this is a consequence of the time the two different strains spend on a pillar because of tumbling, or lack thereof. The extent of $\alpha$ will be determined by steric and hydrodynamic interactions. Its value could be predicted theoretically and compared with our values to establish the relative importance of these interactions.

We have shown that the overall transport of both wild-type and smooth-swimming mutant was diffusive, even in very dense obstacle lattices, where diffusivity retained high values, as predicted by \cite{Jakuszeit2019}. Our experimental values for the diffusivity were compared with theoretical values from an RTP model.
As we did not observe any trapping in the lattice of regular obstacles, we were able to apply a modified run-and-tumble model to estimate the effective diffusion coefficient for each bacterial strain. In free space, it is well known that tumbling decreases the diffusion coefficient $D_0=v l_p$ due to the decrease in persistence length $l_p$. However, the effect of tumbling in more complex environments remains an open question.
While the frequency of tumbling was reduced at a flat wall \cite{Molaei2014}, there have long been suggestions that tumbling can be a means to escape dead-ends in more complex environments, see \cite{Wolfe1989, Bhattacharjee2019, Irani2022}.
Our results in the simpler obstacle lattice system show that tumbling can facilitate escape from surfaces, most notably by reducing the residence time, but might also itself be altered by surface effects. 

To conclude, we have shown that bacterial scattering off cylindrical pillars is non-trivial, particularly when swimmers can tumble, as most wild-type bacteria do. The microscopic dynamics of how swimmers scatter and are trapped by a porous matrix are critical to determining their transport, e.g. through soil \cite{Knights2021, Ford2007} or when infecting tissue \cite{Lux2001, Ribet2015, Balzan2007}.  In future it will be interesting to adapt our analysis of scattering in regular obstacle lattices to more complex porous environments. This will allow to test theoretical predictions for bacteria in porous media from 2D \cite{Saintillan2023} and 3D \cite{Bhattacharjee2019} models.



\section*{Acknowledgements}
We acknowledge financial support from EPSRC EP/L504920/1 and EP/N509620/1 (T.J.) and the Winton Programme for the Physics of Sustainability (T.J., O.A.C.). T. J. also acknowledges support from HFSP. We thank Jana Schwarz-Linek and Angela Dawson for kindly providing the bacterial strains used in this study. We thank Will Arter and Karem Al Nahas for assistance with the microfabrication, and Samuel Bell for illuminating discussions.

\appendix

\section{Effective swimming speed}

We follow the definition for the effective swimming speed given in \cite{Jakuszeit2019} for a sliding boundary condition. Assuming a particle that follows run-and-tumble dynamics, we introduce the mean run time $\tau$. In an obstacle lattice, $\tau$ is a combination of the time between obstacle collisions $\tau_c$ and the time spent at an obstacle $\tau_R$, i.e. $\tau=\tau_c+\tau_R$. The mean time between collisions depends purely on the level on confinement and is thus $\tau_c=\lambda/v$, where $\lambda$ is the mean free path given by Santalo's forumla and $v$ is the swimming speed. For a fixed polar angle $\alpha$, the mean residence time at the obstacle follows as $\tau_R=R \alpha/v$, where $R$ is the obstacle radius.
A cell that travels around a convex obstacle covers a distance $l< v \, \tau_R$. We thus obtain $v_{obs} = l/\tau_R$. The distance $l$ follows from the cosine rule as $l = R  \sqrt{2-2\cos \alpha}$. The effective speed is then the average of the speed at an obstacle and the speed in free space, i.e. 
\begin{equation}
\tilde{v} = v_{obs} \frac{\tau_R}{\tau} + v \frac{\tau_c}{\tau} = \frac{l}{\tau}+v\frac{\tau_{c}}{\tau}.
\label{Eq: effective speed}
\end{equation}

\bibliographystyle{apsrev4-2} 
\bibliography{references}

\end{document}